\documentclass[a4paper]{jpconf}
\usepackage{graphicx}
\begin{document}
\title{A Short Review of  the $S_4$ Symmetric Microscopic Model  for Iron-Based High Temperature Superconductors}

\author{Jiangping Hu}

\address{Beijing National
Laboratory for Condensed Matter Physics, Institute of Physics,
Chinese Academy of Sciences, Beijing 100080,
China  \&  Department of Physics, Purdue University, West
Lafayette, Indiana 47907, USA} 

\ead{jphu@iphy.ac.cn \& hu4@purdue.edu}

\begin{abstract}
We briefly review  the recently constructed two orbital microscopic model   for iron-based superconductors based on $S_4$ symmetry\cite{Hu2012s4}.   With  this faithful representation of the kinematics of the tri-layer FeAs or FeSe structure,  the model provides   answers and physical pictures to  fundamental questions related to the robustness of superconductivity and pairing symmetry, unifies  different families of iron-based superconductors, casts new insight into the connections to the other high $T_c$ superconductors, cuprates, and reveals why an s-wave pairing can be stabilized by repulsive interactions.  Further progresses  include that the model provides a clean understanding of band reconstruction observed in magnetically ordered states, which is a strong support to  the kinematics  of the $S_4$ model,  and    captures the essential low energy physics of iron-based superconductors based on   numerical results from unbiased quantum Monte Carlo simulation.

\end{abstract}

\section{Introduction}
In the past four years, the main theoretical difficulty in  understanding iron-based high temperature superconductors\cite{Hosono,ChenXH,wangnl2008,ChenXL}   stems from the fact that the materials are intrinsic multi-orbital systems due to the presence of all five  3d-orbitals of iron atoms.  Several initial attempts to construct simplified models with fewer orbitals was made in an early stage of the research\cite{daghofer2010,leewen2008,raghu2008}. While these  models could successfully explain some properties, they all had serious drawbacks, which resulted in a belief that  any microscopic model composed of less than all five  $d$-orbitals  is insufficient\cite{hirschfeld,kuroki}.

Recently,  very surprisingly,  with a proper consideration of  the S$_4$ symmetry, the symmetry of  the building block--the   FeAs or FeSe trilayer, we demonstrate that an effective two-orbital model is sufficient to capture the kinematics responsible for low energy physics \cite{Hu2012s4}.   The two orbital model includes two nearly degenerate and weakly coupled single-orbital parts that can be mapped to each other under the $S_4$ transformation.   Not only has the model greatly simplified kinematics which  essentially can be described  by only  three important tight-binding hopping parameters, but the model also has manifested  a clear physical picture   to explore many intriguing fundamental properties of the materials and has drawn a deep connection to the other high temperature superconductors, cuprates.  In this paper, we give a quick review of the model  by emphasizing its unique aspects and summarizing recent numerical and analytic results\cite{mahu2012,haohu2012}.

\section{Motivation}
Before  proceeding to the S$_4$ model,  we  discuss  several main initial motivations which lead to the construction of the model.
The motivations stem from  a few intriguing questions related to iron-based superconductors. 
\begin{itemize}
\item {\it What is the relation between iron-based superconductors and cuprates?} This question has undoubtedly been asked by many researchers   since iron-pnictides were discovered.  In the past several years, it has been shown that the two classes of materials share many similar properties, including similar  layered lattice structures, phase diagrams,  strong antiferromagnetism in parental compounds, short superconducting coherent lengths and so on. However, it is still highly controversial that whether iron-based superconductors allow us to finally apply  the induction method for high temperature superconductors. 

\item{\it  How can the two different families of iron-based superconductors, iron-pnictides and iron chalcogenides, be unified? }
Iron-based superconductors include two families, iron-pnictides\cite{Hosono,ChenXH,wangnl2008} and iron-chalcogenides\cite{ChenXL}.  They share many intriguing common properties. They both have the highest $T_c$s   around 50K\cite{john,ChenXH,liud,qwang,lsun}. The superconducting gaps  are close to isotropic   around Fermi surfaces\cite{hding, zhouxj, zhangy2,Wang_122Se, Zhang_122Se,Mou_122Se}  and the ratio between the gap  and $T_c$, $2\Delta/T_c$,   are much larger than the BCS ratio, 3.52, in both families.  However, the electronic structures in the two  families, in particular, the Fermi surface topologies, are quite different in the materials  reaching  the highest $T_c$.   The hole pockets are absent in iron-chalcogenides but present in iron-pnictides\cite{hding,Wang_122Se, Zhang_122Se,Mou_122Se}. The presence of the hole pockets has been a necessity for superconductivity in  the majority of studies and models which deeply depend on the properties of  Fermi surfaces\cite{hirschfeld,john,Dongj2008,Mazin2008,Kuroki2011, WangF, thomale1, thomale2, chubukov,zlako}. Therefore, the absence of the hole pockets in iron-chalcogenides causes a strong debate over whether both families belong to the same category that shares a common superconducting mechanism.  

\item{\it Why and how can a s-wave pairing state be stabilized if  the superconductivity is originated from a repulsive interaction?}
With a d-wave pairing symmetry, such as in cuprates,  the positive and negative signs of the superconducting order parameter  are equally distributed in both real and reciprocal  space .   Therefore, one can argue that the onsite repulsive interaction is naturally avoided in a d-wave pairing state.  For a s-wave pairing symmetry, even in a sign changed $s^\pm$ state\cite{Mazin2008},  the argument can not be held in general. 
While many of us use repulsive interactions to obtain  a s-wave pairing or $s^\pm$,  in iron-pnictides, this fundamental problem is largely  unaddressed theoretically.

\item  {\it Why is the superconductivity  so robust and how can a single energy scale dominate in pairing gap functions  in iron-based superconductors?} Observed by angle-resolved photoemission microscopy (ARPES),  a very intriguing property in the SC states of iron-pnictides is that the SC gaps around optimal doping on different Fermi surfaces are nearly proportional to a simple form factor $cos k_x cosk_y$ in reciprocal space. This form factor has been observed in both  $122$\cite{hding, hding2, zhouxj,Nakayama}  and $111$\cite{umezawa,liuzh} families of iron-pnictides.  Just like the $d$-wave form factor $cosk_x-cosk_y$ in cuprates, such a form factor indicates that  the pairing between two  next nearest neighbour (NNN) iron sites in  real space dominates.  In a multi-orbital model, many theoretical calculations based on weak coupling approaches have shown that the gap functions are very sensitive to detailed band structures and  vary significantly when  the doping changes\cite{hirschfeld, WangF, thomale1, thomale2, chubukov,zlako}. The robustness of the form factor has been  argued to favor strong coupling approaches which emphasize  electron-electron correlation or the effective  next nearest neighbour (NNN) antiferromagnetic (AF) exchange coupling $J_2$\cite{seo2008,Fang2011,local1, hu1,hu4,luxl,berg,bergkivelson2010} as a primary source of  the pairing force.   
However, realistically, it is very difficult to imagine there is a single energy scale in local exchange interactions if  a multi $d$-orbital model is considered.  
\end{itemize}
In order to answer the above questions, it is clear that  we have to identify  microscopic parameters that control the essential physics of iron-based superconductors at low energy.

\section{Gauge mapping between the symmetries of  superconductivity pairing and  kinematics}
The idea to address the robustness of possible s-wave pairing in iron-based superconductors stems from a paper by Berg, Scalapino and Kivelson\cite{bergkivelson2010}, where they showed that the s-wave pairing in a two leg-ladder model for iron-pnictides is equivalent to a d-wave pairing in a similar model of cuprates.  The s-wave and the d-wave can be  mapped by a simple gauge transformation.  Although their idea was carried out in an one-dimensional model,  they raised an indispensable general point for any dimensional model, that is, \emph{the pairing symmetry and kinematics must be considered together}.  In particular, in a system where the short range pairing dominates, the phases or the signs of superconducting order parameters can be exchanged with those of the local hopping parameters.

\subsection{Gauge Mapping}
A gauge mapping can be generally described as follows.   Given a general Hamiltonian including a kinetic energy  $\hat H_0$ and an  interaction energy  $\hat H_I$,  we can perform  a gauge  mapping that changes the sign of local Fermionic operators, namely,  
\begin{equation}
c_i\rightarrow (-1)^{\theta_i}c_i
\end{equation} where $\theta_i=0,$ or $1$.  It is clear that such a general gauge mapping can be described by an unitary operator $\hat U$. Under the mapping,  the hopping terms in $\hat H_0$ can change the sign so that  the kinetic energy Hamiltonian is changed. However, the eigenvalues of $\hat H_0$   are not changed since it is an unitary transformation. It is also easy to see that the interaction  Hamiltonian $\hat H_I$,  if it includes standard interaction terms, such as local density interaction, spin exchange interaction and so on, in general, is not changed under the gauge transformation. Therefore, under the mapping, 
\begin{equation}
\hat H= \hat H_0+\hat H_I \rightarrow \hat H'= \hat U^+\hat H\hat U=  \hat H'_0+\hat H_I. 
\end{equation}
 If the original Hamiltonian $\hat H$ has a superconducting ground state with  an superconducting order parameter
 $<c_ic_j>=\Delta_{ij}$, the ground state of the new Hamiltonian $\hat H'$  must have $<c_ic_j>=(-1)^{\theta_i+\theta_j}\Delta_{ij}$. Therefore, if the pairing is rather local short range, it is possible to change order parameter symmetry by choosing a proper gauge mapping. 
 
 \subsection{Gauge   mapping between  d-wave and extended s-wave pairing symmetries in cuprates}
\begin{figure}[tbp]
\begin{center}
\includegraphics[width=0.8\linewidth]{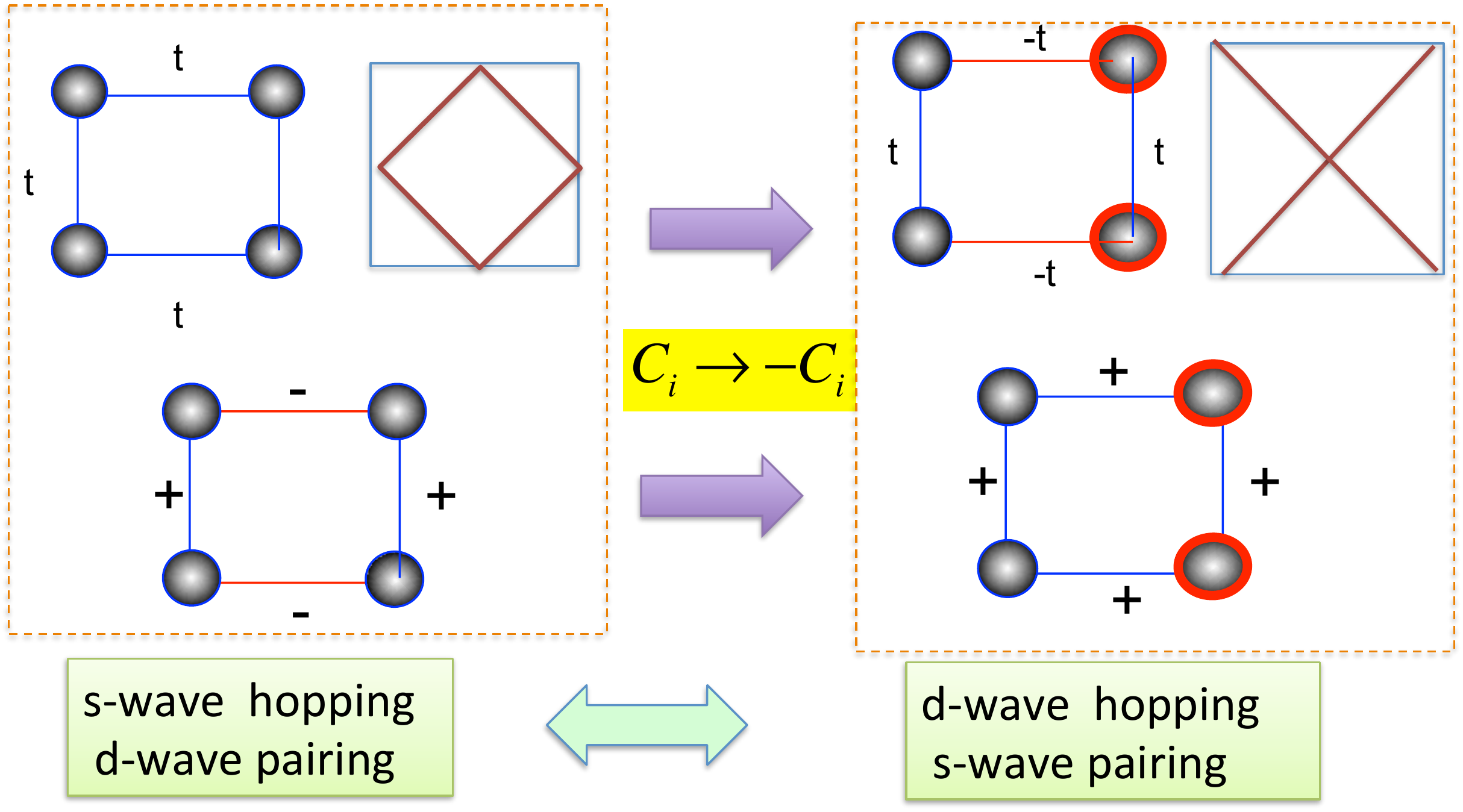}
\end{center}
\caption{ An illustration of the gauge mapping for cuprates. The left part of the figure shows the symmetry of the hopping parameters, d-wave pairing symemtry and the  simplified Fermi surfaces at half filling. The right part   shows their corresponding parts after the gauge mapping. The gague mapping changes the sign of Fermionic operators at the red sites. }\label{cuprates}
\end{figure}

Let's use cuprates as an example to illustrate the point.   For cuprates, the kinetic energy $\hat H_0$ is dominated by the nearest neighbour(NN) hopping in a square lattice, namely, 
\begin{equation}
H_0=-t\sum_{<ij>}c_{i\alpha}^+c_{j\alpha} +...
\end{equation} 
where $<ij>$ represents the links between two NN sites,  and the order parameter of the d-wave pairing symmetry in real space is also dominated by coupling two NN sites which is corresponding to the  d-wave gap form $cosk_x-cosk_y$ in  reciprocal space, namely,
\begin{equation}
<c_ic_{i+\vec y}>= - <c_ic_{i+\vec x}>=\Delta_0
\end{equation} 
where $\vec x,\vec y$ are lattice unit vectors along x and y directions respectively.   As shown in fig.\ref{cuprates},   if we set a gauge mapping as \begin{equation}
c_{i\alpha}\rightarrow (-1)^{i_x}c_{i\alpha},
\end{equation} where $i_x$  labels the coordinate along the x direction  of the ith site. After the mapping, we  have  
\begin{equation}
H'_0=-t\sum_{<ij>_y}c_{i\alpha}^+c_{j\alpha} + t\sum_{<ij>_x}c_{i\alpha}^+c_{j\alpha} ...
\end{equation} 
and 
\begin{equation}
<c_ic_{i+\vec y}>=  <c_ic_{i+\vec x}>=\Delta_0
\end{equation} 
Therefore,  the gauge mapping allows us to exchange the symmetries between the hopping parameters and the superconducting pairing order parameters.  Before the gauge mapping, the hopping has a s-wave symmetry and the pairing has a d-wave symmetry. After the gauge mapping, the hopping has a d-wave symmetry and the pairing has a s-wave symmetry.
It is also important to notice that the Fermi surface topology is changed after the gauge mapping as shown in fig.\ref{cuprates}. In a weak coupling approach, the Fermi surface topology has been the key ingredient to determine many properties, such as pairing symmetry. If the pairing is driven by repulsive interaction  or antiferromagnetic fluctuations, the Fermi surface topology before the gauge mapping favors a d-wave symmetry while one can argue that the Fermi surface topology after the gauge mapping favors a s-wave symmetry.  Therefore, the result obtained from a simple Fermi surface topology analysis is actually consistent with the gauge mapping result.  We will revisit this point for iron-based superconductors in the following subsection.

\subsection{Gauge mapping between  extended s-wave  and d-wave pairing symmetries in iron-based superconductors}
\begin{figure}[tbp]
\begin{center}
\includegraphics[width=0.8\linewidth]{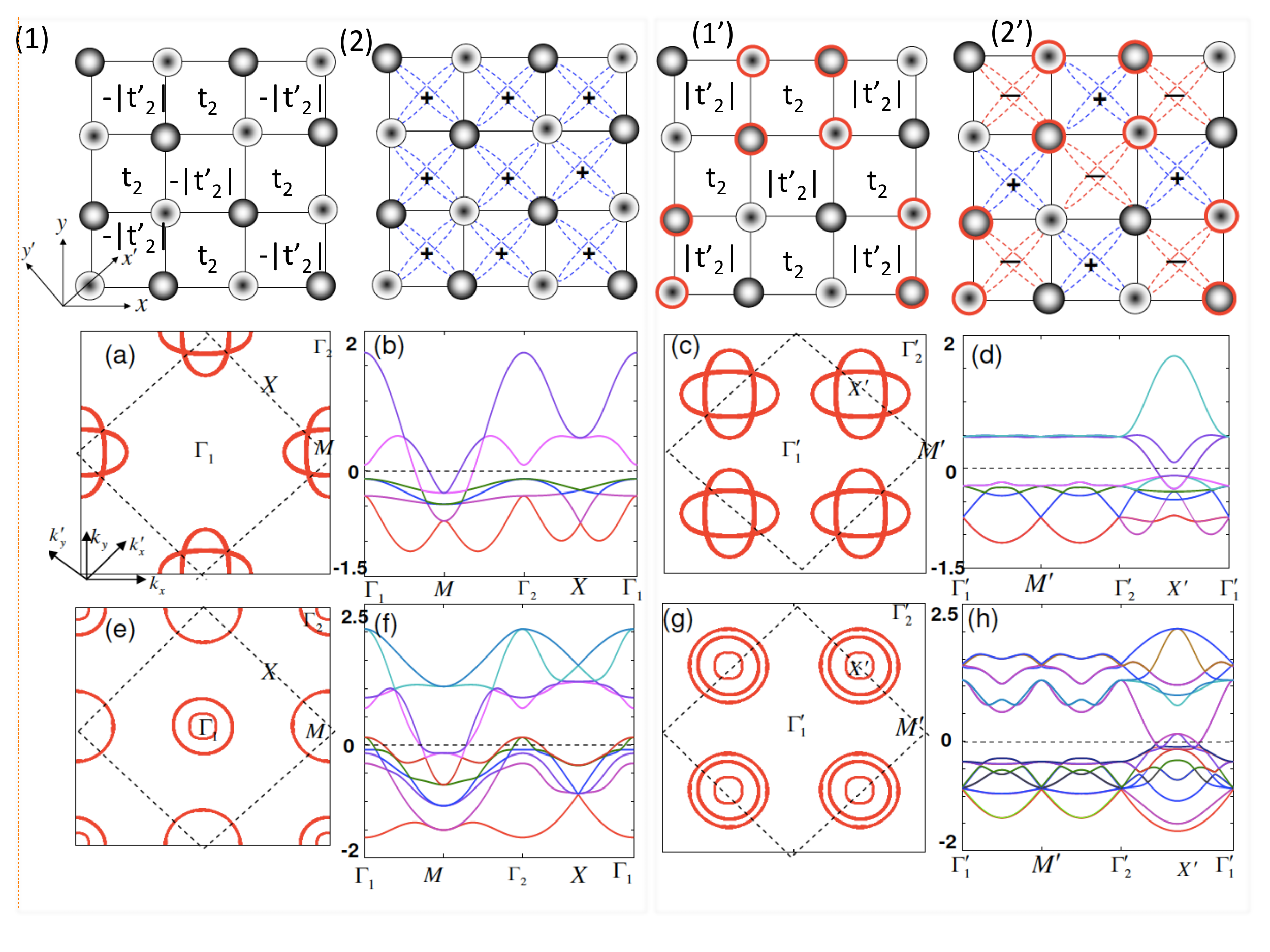}
\end{center}
\caption{  The illustration of the gauge mapping for iron-based superconductors: (1,2) and ($1',2'$) sketch hopping parameters and pairing symmetries before and after the gauge mapping respectively; (a-h) band structures and Fermi surfaces  for two models\cite{Fang2011,kuroki}  before and after the gauge mapping: (a,e)  the Fermi surfaces, (b,f) the band dispersion along the high symmetry lines, (c,g) the Fermi surfaces after the gauge transformation, (d,h) the band dispersions along the high symmetry lines after the gauge transformation.  The  hopping parameters can be found in the above two references. }\label{iron-mappinga}
\end{figure}

 The above analysis has been generalized to iron-based superconductors\cite{Hu2012s4}. A similar  gauge mapping  which maps a   s-wave  to a  d-wave pairing symmetry  can be designed as shown in fig.\ref{iron-mappinga}$(1,2,1',2')$. We group four neighbour iron sites to form a super site  and change the sign of Fermionic operators at the red iron sites  in fig.\ref{iron-mappinga}$(1')$.  
This transformation maps the $A_{1g}$ $s$-wave $cos(k_x)cos(k_y)$ pairing symmetry in the original Fe lattice to a familiar $d$-wave $cosk'_x -cosk_y'$ pairing symmetry defined in the two sublattices of the orginal lattice, where $(k_x,k_y)$ and $(k_x',k_y')$ label momentum in Brillouin zones of the origin lattice and  sublattice respectively.  This is actually a straightforward generalization of the transformation used  in the study of the one dimensional  iron ladder model\cite{bergkivelson2010}.
  
Following the same procedure, we can ask how  the kinetic energy changes. In ref.\cite{Hu2012s4}, we have shown that all the different tight binding models constructed for iron-based superconductors, after the gauge mapping, display very striking simplicity: (1)   all Fermi surfaces after the gauge mapping  are relocated around $X'$,  the  anti-nodal points in  a standard $d$-wave superconducting state in the sublattice Brillouin zone; (2)   the bands previously located at the different places  are magically linked in the new gauge setting.  In particular,    the two bands that contribute to electron pockets are nearly degenerate and the bands that contribute to hole pockets are remarkably connected to the former ones.

\begin{figure}[tbp]
\begin{center}
\includegraphics[width=0.8\linewidth]{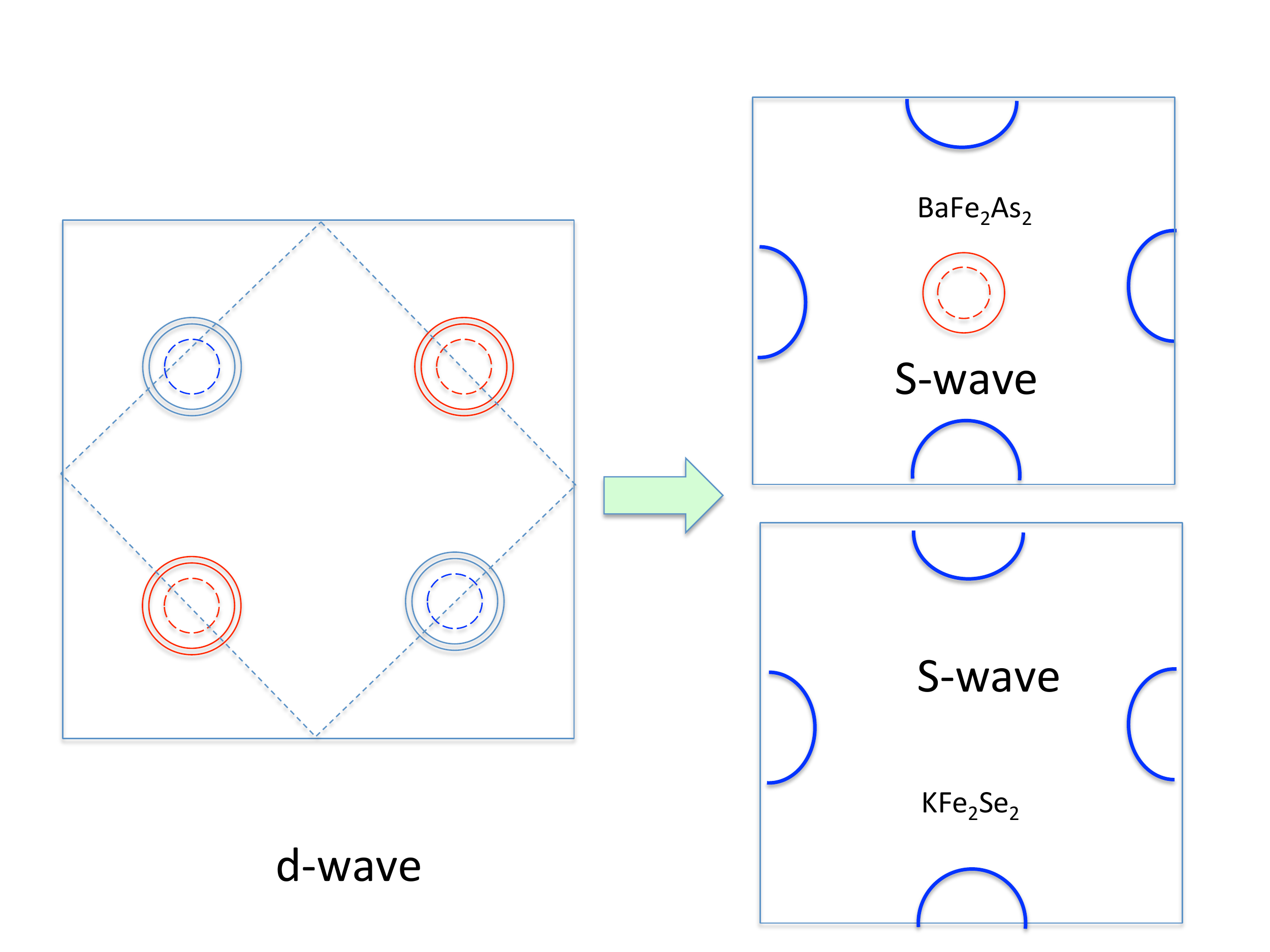}
\end{center}
\caption{ A sketch of a reverse gauge mapping from d-wave to s-wave for both iron-pnictides and iron-chalcogenides.}\label{reverse}
\end{figure}

 The first feature seriously challenges  the consistency of previous weak coupling analysis depending on  the difference of Fermi surface topologies for iron-pnictides and iron-chalcogenides. After the gauge mapping, based on the new  Fermi surface topology, a standard weak coupling analysis would naturally  favor  a robust $d$-wave superconducting state with respect to  the sublattice  in the presence of repulsive interaction or NN AF coupling\cite{scalapino,hu1}. Therefore,  as shown in fig.\ref{reverse}, if we reversely map to the original gauge setting,  the original Hamiltonian must have a robust $s$-wave pairing symmetry. Namely, based on the new Fermi surface topology,  both iron-pnictides and iron-chalcogenides should have  $s$-wave pairing symmetry before the gauge mapping  as long as  the Fermi pockets are small.

The second feature   is the key  feature that leads us to believe that   the low energy physics  is controlled by a two orbital model which is characterized by two nearly degenerate bands.  The  emergence of  the universal kinematics  in different models suggest that the underlining electronic dynamics should be controlled by a much simplified set of parameters.

\section{The construction of a two-orbital model with the $S_4$ symmetry}
With above observations and  the following important facts in the tri-layer FeAs or FeSe structure,
\begin{itemize}
\item   The $d$-orbitals that form the bands near the Fermi surfaces are strongly hybridized with the p-orbitals of As(Se). Since the $d_{x'z}$ and $d_{y'z}$ have the largest overlap with the $p_{x'}$ and $p_{y'}$ orbitals, it is natural for us to use $d_{x'z}$ and $d_{y'z}$ to construct the model;
\item  The local lattice symmetry at an iron site is $D_{2d}$ in which $S_4$ symmetry is the key symmetry operation;

\item  The  two As(Se) planes are separated in space along c-axis in the tri-layer structure. Since there is little coupling between the $p$ orbitals of the two planes and the hopping integrals  through the p-orbitals  are expected to dominate over the direct exchange couplings between the $d$-orbitals themselves, the two orbital model essentially could 
be decoupled into two nearly degenerate one orbital models;

\item  The decoupling must be controlled by the $S_4$ symmetry since $S_4$ symmetry links  two As(Se) planes;
\end{itemize}   
we draw a  natural conclusion that the two-orbital model should be classified according to the  $S_4$ symmetry and each of the two nearly degenerate one-orbital models should have a unit cell with two iron sites.  
\begin{figure}[tbp]
\begin{center}
\includegraphics[width=0.8\linewidth]{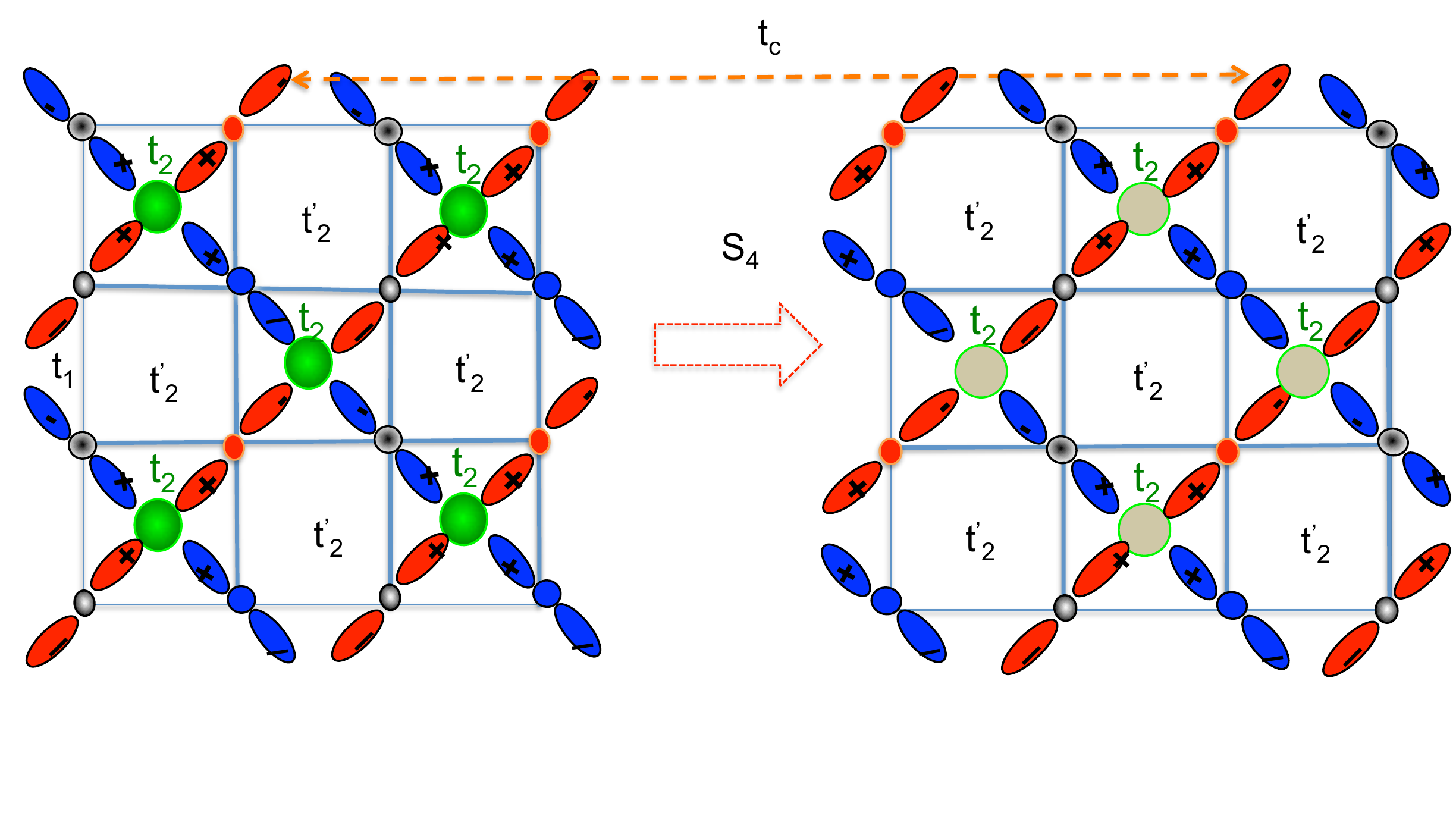}
\end{center}
\caption{ A sketch of the $S_4$ kinematics with $d_{x'z}$ and $ d_{y'z}$ orbitals. The green balls represent As(Se) in the top layer and the grey balls represent those in the bottom layer. The blue and red orbitals represent the $d_{x'z}$ and $d_{y'z}$ orbitals. }\label{s4}
\end{figure}
A picture of the effective two-orbital model with the $S_4$ symmetry is illustrated in fig.\ref{s4}.
The two orbital model is divided into two weakly and coupled  one-orbital models. The orbital degree of freedom in  the two one-orbital models can be called as two components of an $S_4$ iso-spin.      The first one   includes the $d_{x'z}$  in the A sublattice and the $d_{y'z}$ in the B sublattice if we use $A$ and $B$ to label two sublattices, and the second one  includes the $d_{x'z}$  in the B sublattice and the $d_{y'z}$ in the A sublattice.  The first  strongly couples to the top As(Se) layer and  the second couples to  the bottom As(Se) layer. We denote $\hat c_{i\sigma}$ to   $\hat d_{i\sigma}$ to label Fermionic operators of the two models that  are connected by  the $S_4$ transformation.  Then, a general  Hamiltonian that describes the kinematics   is given by, 
\begin{eqnarray}
\hat H_{0,S_4}= \hat H_{0}+\hat H_{0,c},
\end{eqnarray}
where $H_0$ includes the two decoupled one-orbital models and $H_{0.c}$ which also  obey the $S_4$ symmetry describes the weak coupling between the two models. As shown in\cite{Hu2012s4}, if we keep the leading  hopping parameters,
\begin{eqnarray}
  \hat H_{0} & =& \sum_{k\sigma} [4t_{2s} cosk_xcosk_y -\mu](\hat c^+_{k\sigma}\hat c_{k\sigma}+\hat d^+_{k\sigma}\hat d_{k\sigma})\nonumber \\
& & +2t_{1}(cosk_x+cosk_y)(\hat c^+_{k\sigma}\hat c_{k\sigma}-\hat d^+_{k\sigma}\hat d_{k\sigma}) \nonumber \\
& & +2t_{1d}(cosk_x-cosk_y)(\hat c^+_{k\sigma}\hat c_{k\sigma}+\hat d^+_{k\sigma}\hat d_{k\sigma}) \nonumber \\
& &+4t_{2d} sink_xsink_y (\hat c^+_{k\sigma}\hat c_{k+Q\sigma}-\hat d^+_{k\sigma}\hat d_{k+Q\sigma}) \nonumber \\
& &+2t_{3}(cos2k_x+cos2k_y)(\hat c^+_{k\sigma}\hat c_{k\sigma}+\hat d^+_{k\sigma}\hat d_{k\sigma}) \nonumber \\
& & +... 
\label{hh}
\end{eqnarray}
and 
\begin{eqnarray}
\hat H_{0,c}=\sum_k 2 t_c (cosk_x+cosk_y)(\hat c^+_{k\sigma}\hat d_{k\sigma}+h.c.).
\label{tc}
\end{eqnarray} 
where  we limit  the hopping parameters up to the third NN (TNN), including NN hoppings, $t_{1x}$, $t_{1y}$,   NNN hoppings,  $t_2$, $t_2'$, and TNN hoppings, $t_{3}$. The longer range hoppings can be included  if needed.  For convenience, we have  defined $t_{1s}=(t_{1x}+t_{1y})/2$, $t_{1d}=(t_{1x}-t_{1y})/2$, $t_{2s}=(t_2+t'_2)/2$ and $t_{2d}=(t_2-t_2')/2$.  
\begin{figure}[tbp]
\begin{center}
\includegraphics[width=1.0\linewidth]{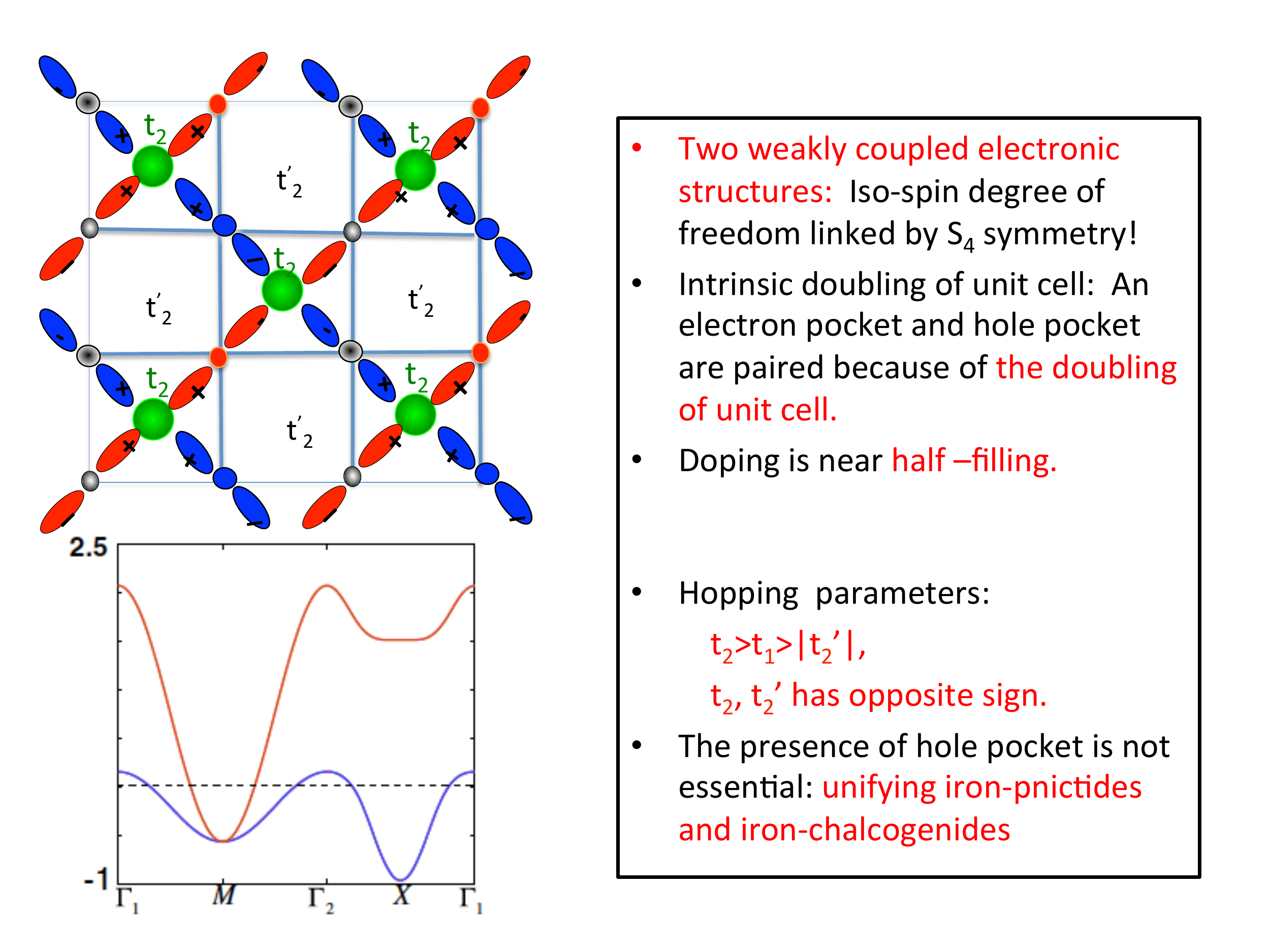}
\end{center}
\caption{ A sketch of the underlining kinematics  and band structure of the $S_4$ model,  and a list of the unique and important properties of the model. }\label{s4model}
\end{figure}


 Ignoring $t_c$,  each $S_4$ iso-spin components  provides the following two paired band energy dispersions due to unit cell doubling,
 \begin{eqnarray}
& & E_{e}=\epsilon_k +
    4 \sqrt{t_{2d} ^2 sin^2x sin^2y  + [\frac{t_{1s} (cosk_x + cosk_y) \pm t_{1d} (cosk_x - cosk_y)}{2}]^2},\\
& & E_{h}=\epsilon_k 
-    4 \sqrt{t_{2d} ^2 sin^2x sin^2y  + [\frac{t_{1s} (cosk_x + cosk_y) \pm t_{1d} (cosk_x - cosk_y)}{2}]^2},
\label{dis}
\end{eqnarray}
 
where $\epsilon_k=4 t_{2s} cosk_x cosk_y + 2 t_{3} (cos2k_x + cos2k_y) - \mu$.
 $E_{e}$ can capture the electron pocket at M points and  $E_{h}$ can capture the hole pocket at $\Gamma$ point.  $t_{1s}$, $t_{2s}$ and  $t_{2d}$  are  the key parameters.
By just keeping these three parameters, the model is already good enough to capture the main characters of the bands contributing to the Fermi surface.  Including interactions, the general Hamiltonian is given by
\begin{eqnarray}
 H=H_{0,s_4}+U\sum_{i,\alpha=1,2} \hat n_{i,\alpha\uparrow}\hat n_{i,\alpha\downarrow}+U'\sum_{i}\hat n_{i,1}\hat n_{i,2}+J_H'\sum_i \hat S_{i,1}\cdot \hat S_{i,2}
\end{eqnarray}
where $\alpha=1,2$ labels the $S_4$ iso-spin,  $U$ describes the effective Hubbard repulsion interaction within each iso-spin, $U'$ describes the one between them and $J'_H$ describes the effective Hund$'$s coupling. Since the two components couple weakly, we may expect $U$ dominates over $U'$ and $J_H'$.  Then, in  the first order  approximation, the model could become a  single orbital-Hubbard model near half-filling. 

 In the original paper\cite{Hu2012s4}, we have suggested that $t_{1s}$ breaks $D_{2d}$ symmetry.  This statement is not absolutely  correct  because $t_{1s}$ and $t_{1d}$ can be exchanged by choosing  different orbital configurations.  With the orbital configuration shown in  fig.\ref{s4}, $t_{1s}$ does not break $D_{2d}$ symmetry.   It is also important to note that the $\Gamma$ point and the $(\pi,\pi)$ point in Brillouin zone here are  exchangeable because of the doubling of the unit cell.

\section{Numerical results and supporting experimental evidence for the $S_4$ model }
In fig.\ref{s4model},  we summarize the important properties in the $S_4$ model. These properties include
\begin{itemize}
\item  For each $S_4$ iso-spin,  there is a  pair of  bands  generated by unit cell doubling: one contributes a hole pocket and the other contributes  an electron pocket.
\item For each $S_4$ iso-spin, the filling factor is close to half-filling, an important property to allow strong spin fluctuation.
\item  The unit cell is intrinsically doubled, which is a very different starting point of kinematics from previous models.
\item  The hole pockets can be suppressed by adjusting $t_{1s}$ or $t_{2s}$. Fixing $t_{2d}$ and $t_{2s}$, reducing $t_{1s}$ leads to flatter dispersion of the band responsible for the hole pockets. Thus, it suppresses the hole pockets. Therefore, the model unifies iron-pnictides and iron-chalcogenides.
\item  From orbital configuration, we have $t_2>|t_{1s}|>|t'_{2}|$ and the signs of $t_2$ and $t'_{2}$ are opposite. 
\end{itemize}
 These characters  have important physical consequences. In the following, we discuss some recent progresses associated with them. 
 \subsection{Quantum Monte Carlo (QMC) Simulation on the $S_4$ model}
The  microscopic understanding  provides a new opportunity to make  use of highly controllable and unbiased numerical methods to study iron-based superconductors.
In a recent paper\cite{mahu2012}, we have demonstrated that the model in the presence of  electron-electron interaction is characterized by strong  $(0,\pi)$ collinear antiferromagentic (C-AFM) fluctuations as observed in iron-based superconductors\cite{spinwave-dela2008, zhao1, DaiHureview}. Moreover,   the pairing with an extended $s$-wave
 symmetry robustly dominates over other pairings at low temperature in reasonable parameter region regardless of the change of Fermi surface topologies, namely, the presence or absence of hole pockets at $\Gamma$ point.  For both iron-pnictides and iron-chalcogenides, the pairing correlation, the effective pairing
interaction and the 
antiferromagnetic  correlation  strongly increase as the on-site Coulomb
interaction increases.  
\begin{figure}[tbp]
\begin{center}
\includegraphics[width=1\linewidth]{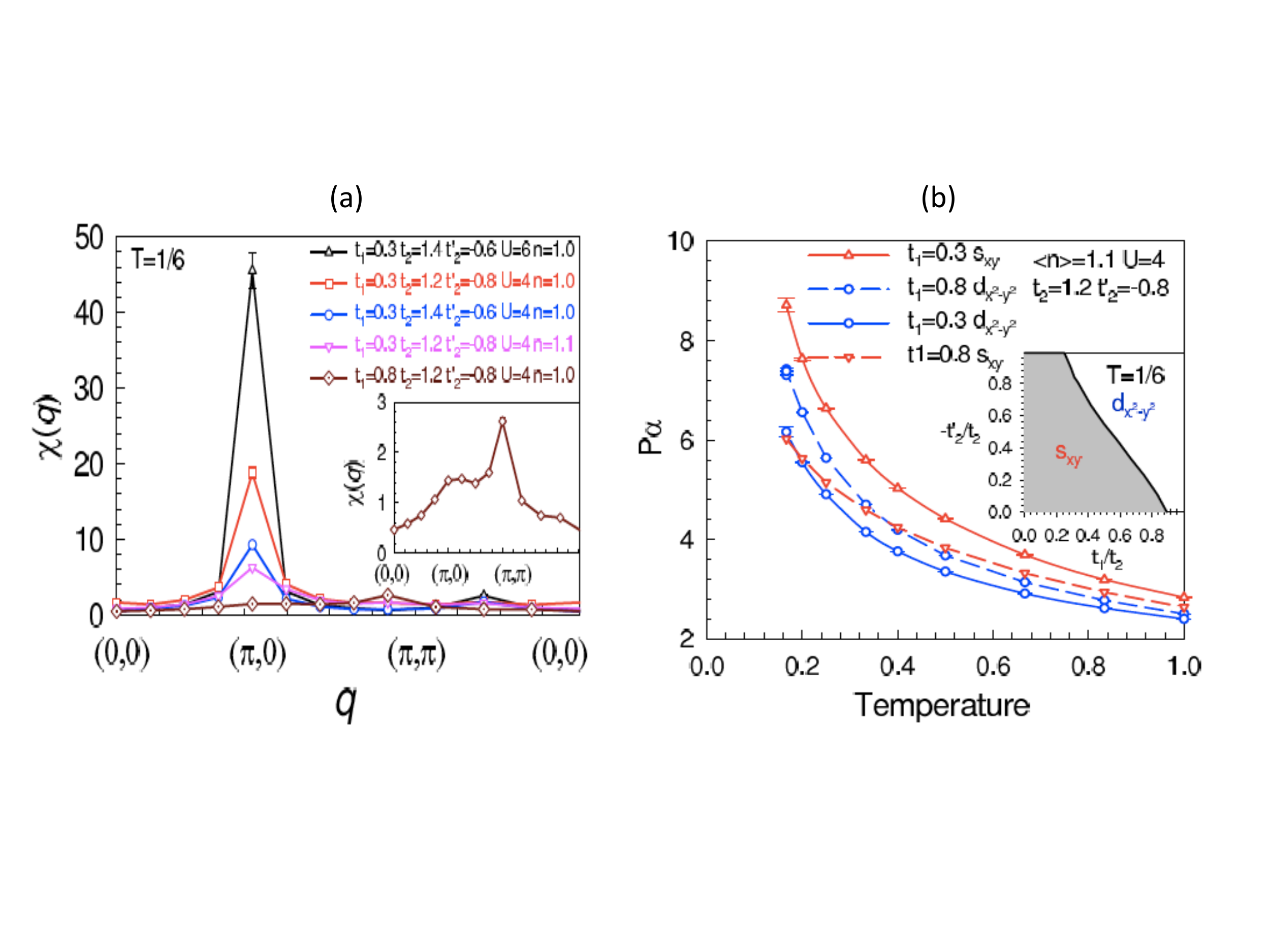}
\end{center}
\caption{  (a)  Quantum Monte Carlo calculations about the static spin susceptibility for the $S_4$ model. The inset shows a significant large  $t_1$ can change  the peak from $(0,\pi)$ to $(\pi,\pi)$, which signals the competition between NN hopping and NNN hopping.  (b) The effective pairing interaction as a function of temperature
at $<n>$=1.1, $U=4.0$, $t_2=1.2$, $t'_2=-0.8$ for different $t_1$.
Inset: the competition between $s_{xy}$ and $d_{x^2-y^2}$ depends on $t_1/t_2$ and $-t'_2/t_2$ .}\label{qmc}
\end{figure}

QMC results are summarized in fig.\ref{qmc}.
The competition between the $s$-wave and the $d$-wave in the $S_4$model is manifested  by $t_1$ and $t_2$ where $t_1$ favors the $d$-wave while $t_2$ favors the $s$-wave. A simple phase diagram from QMC simulation is given as the inset in  fig. \ref{qmc}(b). For iron-based superconductors,  $t_1$ should be roughly equal to $t_2/2$ and $|t'_2|$ is much smaller than $t_1$ and $t_2$. Therefore, the $s$-wave should be robustly favored.  

\subsection{ Oriented gap opening in the C-AFM state: an impact of intrinsic  unit cell doubling}
As we have discussed above, one unique feature in the $S_4$ symmetry construction is that the unit cell includes 2 irons.  Most of previous studies are based on   models with one iron per unit cell. In principle, one can argue that models based on both representations should be equivalent. However, since the search of a meanfield result  depends on  the starting point of  kinematics, a proper starting point  is crucial to obtaining a correct meanfield result.
 \begin{figure}[tbp]
\begin{center}
\includegraphics[width=1\linewidth]{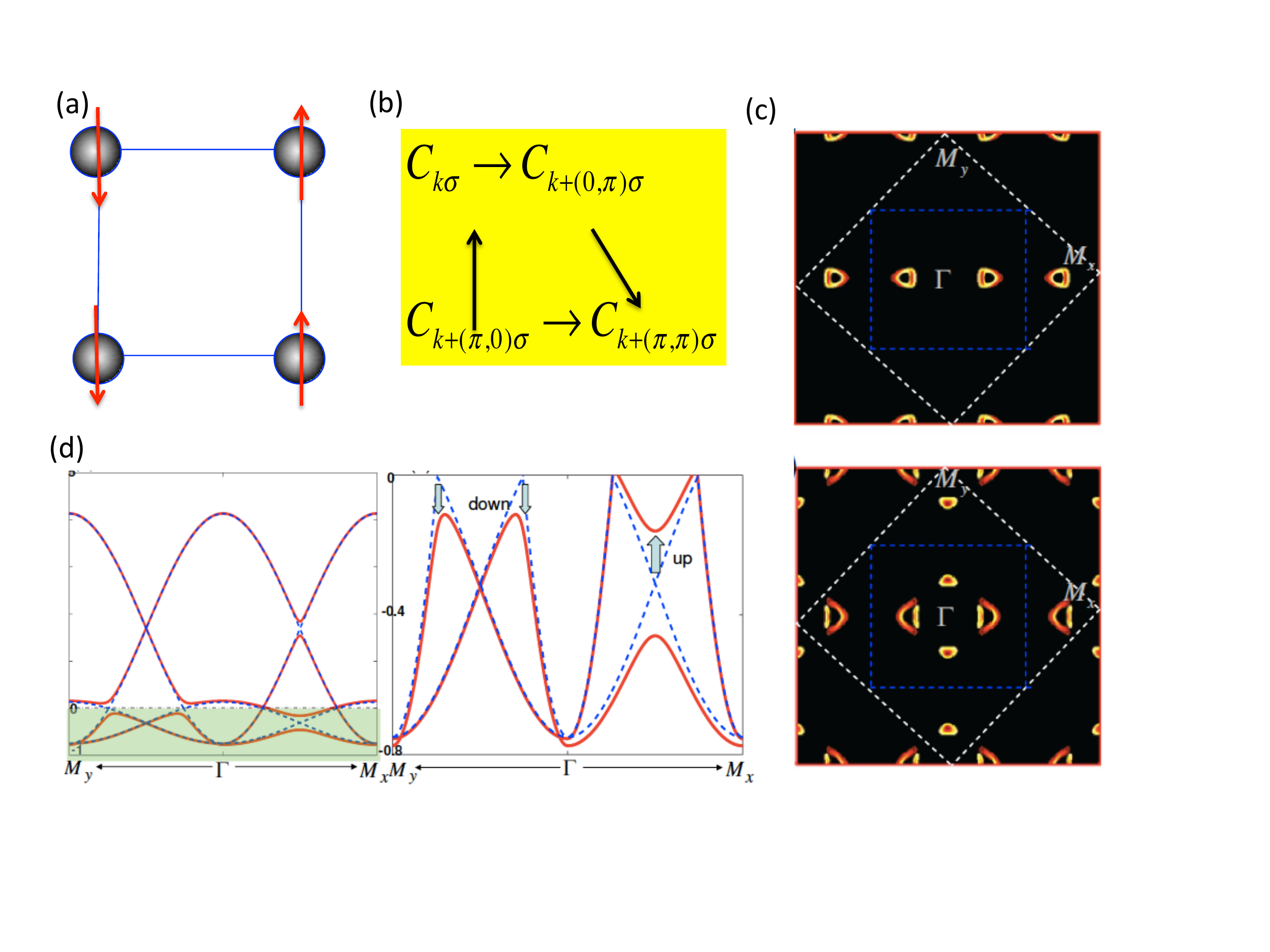}
\end{center}
\caption{The illustration of oriented gap opening in magentically ordered state of iron-based superconductors: (a) The C-AFM magnetic state; (b) The illustration of the connections of four Fermionic operators in reciprocal space by the C-AFM order; (c) The typical Fermi surfaces in the C-AFM state; (d) The band reconstruction and gap opening in the C-AFM state.  }\label{gap}
\end{figure}

In a recent paper, we have demonstrated an important impact of the intrinsic unit cell doubling\cite{haohu2012}.  Unlike in a conventional spin density wave (SDW)  state where gaps at any two points on the Fermi
surface that  are connected by the ordered SDW  wave vector should be opened, the band structure in the C-AFM state appears to be reconstructed by the development
of the magnetic order. ARPES experiments \cite{miyi2009,richardarpes2010,nematic-yi2010,zhangapres2012,zhouxj2008, zhouxj2009} observe that
small pockets or hot spots develop around Fermi surfaces. It is fair to
say that the scenario of the gap opening due to a magnetic order in a
multi-orbital system, such as iron-pnictides, is complex. For example, it has been argued that the entire Fermi surfaces can be gaped out\cite{ranying2009}.  Nevertheless, the surprising band
reconstruction is never fully understood in previous models.

As summarized in fig.\ref{gap}, the complicated band reconstruction near Fermi surfaces in the
magnetically ordered state of iron-pnictides observed by ARPES can be  understood in a
meanfield level  by the  S$_{4}$   model.  The (0,$\pi$) or ($\pi$,0) 
C-AFM order does not open gaps between two points on theFermi
surfaces linked by the ordered wave vector but 
forces a band reconstruction involving four points in unfolded Brillouin zone
 and gives rise to small pockets or hot spots.  

  The doubling of the unit cell leads to  an orientation-dependent gap opening in reciprocal space. In one $\Gamma-M$ direction, the magnetic order splits two folded bands while in the other direction the gap can be opened between the pair of electron and hole pocket bands. Such an oriented gap opening generates the hot spots or small pockets  as observed by ARPES as shown in fig.\ref{gap}(c)..

\subsection{ Sign change between the top and bottom As(Se) planes in the $s$-wave state}
The NNN hopping parameters $t_2$ and $t'_2$ have opposite signs in the $S_4$ model.  This sign difference stems from the orbital symmetries.  If we take the gauge mapping mentioned earlier,  after the mapping, the two hopping parameters in the  new model have the same signs.   Therefore, the gauge mapping  exactly switches the symmetries between the pairing and the underlining kinematics as shown in fig.\ref{iron-mappinga}$(1,2,1',2')$.

In the view of the $d$-wave picture after the gauge mapping,     the superconducting phase  for each $S_4$ iso-spin  alternates between neighboring squares, which is corresponding to the sign change  between the top and bottom As/Se planes  in the view of the $S_4$ symmetry. Such an intrinsic sign change between the top and bottom As planes along c-axis explains why a s-wave before gauge mapping can be stabilized even if the repulsive interaction is responsible for superconductivity.

The relative phase between the two $S_4$ iso-spins can take two different values, 0 or $\pi$, in principle.  The difference can be classified as the irreducible representation of the   $S_4$ symmetry.   The $S_4$ point group has four one-dimensional representations, including $A$, $B $ and $2E$. In the $A$ state, the $S_4$ symmetry is maintained.  In the B state, the state changes sign under the $S_4$ transformation. Therefore,  
  in the A phase,  since the $S_4$ symmetry is not violated, the relative phase between the two iso-spin are equal to $\pi$ in space, while in the B phase, the relative phase   is zero.  
The A phase  in the presence of the hole pockets is an sign-changed s-wave or $s^\pm$ state. Therefore, the model predicts a new superconducting state, B phase state.  However,  it is still an open question whether iron-base superconductors can carry both states or not. 
\section{Discussion and perspectives}
\subsection{Comparison between cuprates and iron-based superconductors} By revealing  the underlining electronic structure,  we can make a close comparison between   cuprates and iron-based superconductors.  In Table.\ref{table}, we list the close relations between two high $T_c$ superconductors. From the table, it is clear that  by determining these physical properties of iron-based superconductors listed in the table can help  to determine  the high $T_c$ superconducting mechanism. 
\begin{center}
\begin{table}[tbp]
\begin{tabular}{|c|c|c||}
\hline properties & iron SCs & cuprates\\ \hline
  pairing symmetry &  $s$-wave & $d$-wave  \\ \hline
underlining hopping symmetry & $d$-wave  & $s$-wave  \\ \hline
dominant pairing form & $cosk_xcosk_y$ & $cosk_x-cosk_y$  \\ \hline
pairing classification symmetry &  $S_4$ & $C_{4v}$  \\ \hline
unit cell& 2 Fe (doubled)  & 1 Cu\\ \hline
AF coupling & NNN $J_2$  & NN $J_1$  \\ \hline
sign change in real space & c-axis  & a-b plane   \\ \hline
filling density  &  half-filling  & half-filling \\ \hline
\end{tabular}
\caption{ A list of the close connections between iron-based superconductors (iron SCs)  and  cuprates.}
\label{table}
\end{table}
\end{center}
\subsection{More supporting evidence for  the $S_4$ symmetry}
The $S_4$ model provides a microscopic understanding to two measured universal behaviors: the  superconducting gap function  that is dominated by the single $cosk_xcosk_y$ form and the almost universal  NNN antiferromagnetic exchange coupling strength $J_2$ measured by neutron scattering\cite{zhao1,lip2,my} .  The existence of such two single energy scale quantities is also a strong support for the $S_4$ model.  If we treat the $S_4$ model in the presence of  strong electron-electron correlation, both $J_2$ and the gap function in the s-wave state  is  determined by the NNN hopping parameters.

The $S_4$ model also suggests that the  properties  of an  iron-based superconductor can be understood by simply observing bands responsible for electron pockets.  Since the physics is dominated by the pair of bands that are responsible for  a pair of hole and electron pockets. The third hole pockets at the $\Gamma$ point observed in iron-pnictides  is not expected to play a significant role.  However, the existence of the third hole pocket may change the doping concentration for   the $S_4$ model in a stoichiometric material. Namely,  in the view of the $S_4$ model, the material could be self electron doped.  For example, if  a large third hole pocket is present, we would expect that   the $S_4$ model could be heavily electron-doped.  This  expectation naturally explains puzzled  physical properties of $LiFeAs$ where a large third hole pocket was observed\cite{putzkelifeas2012}. No static magnetic order is observed in the stoichiometric $LiFeAs$ with $T_c\sim 18K$\cite{tapplifeas2008,wanglifeas2008}.   In the view of the $S_4$ model, the $LiFeAs$ is self electron doped so that many physical properties of the material should be understood from  the heavy electron doped side of  phase diagram\cite{wanglifeas2012}.

\subsection{ Perspectives}
The $S_4$ model provides a fresh starting point to explore novel physics in iron-based superconductors.  Here, we discuss several perspectives for the future research.

First, since  the $S_4$ symmetry stems from the fact  that As stands out of Fe plane,   obvious interesting physics may stem from  c-axis coupling. Recently,  ARPES has provided the existence of  superconducting gap nodes along c-axis in $BaFe_2As_{2-x}P_x$\cite{zhang2012node,qiu2012node}.  It is very likely that the $S_4$ symmetry may  be linked to such a development of nodes since the superconducting phase along c-axis in the $S_4$ model has very intriguing  sign change as we discussed earlier.

Second,   electron nematism observed in iron-pnictides\cite{nematic-fisher2011} can be fully investigated in the $S_4$ model. Namely, the $S_4$ model provides a microscopic base to understand electron nematism.  The $S_4$ symmetry breaking can lead to an electron nematic state.  In the previous theoretical studies,  electron nematism is obtained through phenomenological models\cite{Fang2008nematic, xu2008, Eremin2011a,nematic-fernandes2010}.  

Third, with only a set of three parameters describing kinematics,  it becomes possible to obtain the dependence of these parameters on lattice parameters.   Such an dependence may help us to  explain empirical observations of the connection between the maximum $T_c$s and  the As-Fe-As angles\cite{john}. 

Fourth, it is interesting to investigate  physical properties on surfaces where the $S_4$  symmetry, in principle, is naturally broken.  A natural consequence of such a symmetry breaking  is that   symmetry broken states may be more stable on surfaces than in bulk. 
 
Finally,  the doping phase diagram and  its parameter dependence, as well as many other physical quantities need to be investigated in the $S_4$ model. 

\subsection{Summary}
The $S_4$ model provides great simplicity to study  and unify  the physics of iron-based superconductors. It reveals the deep connections between curpates and iron-based superconductors.  Since the uniqueness of the model stems from the unique structure of FeAs or FeSe tri-layer, revealing  such an underlining microscopic kinematics  cast a new light on the search of   new structures that can produce  high superconducting transition temperatures.

\section{Acknowledgments}
The author acknowledges NN Hao, T. Ma, H. Lin,  Y. Wang for collaborations and  H. Ding, DL Feng, PC Dai, T. Xiang, X. Dai,   S. Kivelson, D. H. Lee,  F. Wang for useful discussions. The work is supported  by the Ministry of Science and Technology of China 973 program(2012CB821400) and NSFC-1190024.

\providecommand{\newblock}{}


\end{document}